\documentclass[letterpaper,compsoc,twoside]{IEEEtran}
\usepackage{cmap} 
\usepackage{ifthen}
\usepackage[T1]{fontenc}
\usepackage[utf8]{inputenc}
\usepackage{amsmath}
\usepackage{color}

\usepackage[font={small,it},labelfont=bf]{caption}
\usepackage{float}

\usepackage{scipy}
\makeatletter
\def\PY@reset{\let\PY@it=\relax \let\PY@bf=\relax%
    \let\PY@ul=\relax \let\PY@tc=\relax%
    \let\PY@bc=\relax \let\PY@ff=\relax}
\def\PY@tok#1{\csname PY@tok@#1\endcsname}
\def\PY@toks#1+{\ifx\relax#1\empty\else%
    \PY@tok{#1}\expandafter\PY@toks\fi}
\def\PY@do#1{\PY@bc{\PY@tc{\PY@ul{%
    \PY@it{\PY@bf{\PY@ff{#1}}}}}}}
\def\PY#1#2{\PY@reset\PY@toks#1+\relax+\PY@do{#2}}

\@namedef{PY@tok@w}{\def\PY@tc##1{\textcolor[rgb]{0.73,0.73,0.73}{##1}}}
\@namedef{PY@tok@c}{\let\PY@it=\textit\def\PY@tc##1{\textcolor[rgb]{0.25,0.50,0.56}{##1}}}
\@namedef{PY@tok@cp}{\def\PY@tc##1{\textcolor[rgb]{0.00,0.44,0.13}{##1}}}
\@namedef{PY@tok@cs}{\def\PY@tc##1{\textcolor[rgb]{0.25,0.50,0.56}{##1}}\def\PY@bc##1{{\setlength{\fboxsep}{0pt}\colorbox[rgb]{1.00,0.94,0.94}{\strut ##1}}}}
\@namedef{PY@tok@k}{\let\PY@bf=\textbf\def\PY@tc##1{\textcolor[rgb]{0.00,0.44,0.13}{##1}}}
\@namedef{PY@tok@kp}{\def\PY@tc##1{\textcolor[rgb]{0.00,0.44,0.13}{##1}}}
\@namedef{PY@tok@kt}{\def\PY@tc##1{\textcolor[rgb]{0.56,0.13,0.00}{##1}}}
\@namedef{PY@tok@o}{\def\PY@tc##1{\textcolor[rgb]{0.40,0.40,0.40}{##1}}}
\@namedef{PY@tok@ow}{\let\PY@bf=\textbf\def\PY@tc##1{\textcolor[rgb]{0.00,0.44,0.13}{##1}}}
\@namedef{PY@tok@nb}{\def\PY@tc##1{\textcolor[rgb]{0.00,0.44,0.13}{##1}}}
\@namedef{PY@tok@nf}{\def\PY@tc##1{\textcolor[rgb]{0.02,0.16,0.49}{##1}}}
\@namedef{PY@tok@nc}{\let\PY@bf=\textbf\def\PY@tc##1{\textcolor[rgb]{0.05,0.52,0.71}{##1}}}
\@namedef{PY@tok@nn}{\let\PY@bf=\textbf\def\PY@tc##1{\textcolor[rgb]{0.05,0.52,0.71}{##1}}}
\@namedef{PY@tok@ne}{\def\PY@tc##1{\textcolor[rgb]{0.00,0.44,0.13}{##1}}}
\@namedef{PY@tok@nv}{\def\PY@tc##1{\textcolor[rgb]{0.73,0.38,0.84}{##1}}}
\@namedef{PY@tok@no}{\def\PY@tc##1{\textcolor[rgb]{0.38,0.68,0.84}{##1}}}
\@namedef{PY@tok@nl}{\let\PY@bf=\textbf\def\PY@tc##1{\textcolor[rgb]{0.00,0.13,0.44}{##1}}}
\@namedef{PY@tok@ni}{\let\PY@bf=\textbf\def\PY@tc##1{\textcolor[rgb]{0.84,0.33,0.22}{##1}}}
\@namedef{PY@tok@na}{\def\PY@tc##1{\textcolor[rgb]{0.25,0.44,0.63}{##1}}}
\@namedef{PY@tok@nt}{\let\PY@bf=\textbf\def\PY@tc##1{\textcolor[rgb]{0.02,0.16,0.45}{##1}}}
\@namedef{PY@tok@nd}{\let\PY@bf=\textbf\def\PY@tc##1{\textcolor[rgb]{0.33,0.33,0.33}{##1}}}
\@namedef{PY@tok@s}{\def\PY@tc##1{\textcolor[rgb]{0.25,0.44,0.63}{##1}}}
\@namedef{PY@tok@sd}{\let\PY@it=\textit\def\PY@tc##1{\textcolor[rgb]{0.25,0.44,0.63}{##1}}}
\@namedef{PY@tok@si}{\let\PY@it=\textit\def\PY@tc##1{\textcolor[rgb]{0.44,0.63,0.82}{##1}}}
\@namedef{PY@tok@se}{\let\PY@bf=\textbf\def\PY@tc##1{\textcolor[rgb]{0.25,0.44,0.63}{##1}}}
\@namedef{PY@tok@sr}{\def\PY@tc##1{\textcolor[rgb]{0.14,0.33,0.53}{##1}}}
\@namedef{PY@tok@ss}{\def\PY@tc##1{\textcolor[rgb]{0.32,0.47,0.09}{##1}}}
\@namedef{PY@tok@sx}{\def\PY@tc##1{\textcolor[rgb]{0.78,0.36,0.04}{##1}}}
\@namedef{PY@tok@m}{\def\PY@tc##1{\textcolor[rgb]{0.13,0.50,0.31}{##1}}}
\@namedef{PY@tok@gh}{\let\PY@bf=\textbf\def\PY@tc##1{\textcolor[rgb]{0.00,0.00,0.50}{##1}}}
\@namedef{PY@tok@gu}{\let\PY@bf=\textbf\def\PY@tc##1{\textcolor[rgb]{0.50,0.00,0.50}{##1}}}
\@namedef{PY@tok@gd}{\def\PY@tc##1{\textcolor[rgb]{0.63,0.00,0.00}{##1}}}
\@namedef{PY@tok@gi}{\def\PY@tc##1{\textcolor[rgb]{0.00,0.63,0.00}{##1}}}
\@namedef{PY@tok@gr}{\def\PY@tc##1{\textcolor[rgb]{1.00,0.00,0.00}{##1}}}
\@namedef{PY@tok@ge}{\let\PY@it=\textit}
\@namedef{PY@tok@gs}{\let\PY@bf=\textbf}
\@namedef{PY@tok@gp}{\let\PY@bf=\textbf\def\PY@tc##1{\textcolor[rgb]{0.78,0.36,0.04}{##1}}}
\@namedef{PY@tok@go}{\def\PY@tc##1{\textcolor[rgb]{0.20,0.20,0.20}{##1}}}
\@namedef{PY@tok@gt}{\def\PY@tc##1{\textcolor[rgb]{0.00,0.27,0.87}{##1}}}
\@namedef{PY@tok@err}{\def\PY@bc##1{{\setlength{\fboxsep}{\string -\fboxrule}\fcolorbox[rgb]{1.00,0.00,0.00}{1,1,1}{\strut ##1}}}}
\@namedef{PY@tok@kc}{\let\PY@bf=\textbf\def\PY@tc##1{\textcolor[rgb]{0.00,0.44,0.13}{##1}}}
\@namedef{PY@tok@kd}{\let\PY@bf=\textbf\def\PY@tc##1{\textcolor[rgb]{0.00,0.44,0.13}{##1}}}
\@namedef{PY@tok@kn}{\let\PY@bf=\textbf\def\PY@tc##1{\textcolor[rgb]{0.00,0.44,0.13}{##1}}}
\@namedef{PY@tok@kr}{\let\PY@bf=\textbf\def\PY@tc##1{\textcolor[rgb]{0.00,0.44,0.13}{##1}}}
\@namedef{PY@tok@bp}{\def\PY@tc##1{\textcolor[rgb]{0.00,0.44,0.13}{##1}}}
\@namedef{PY@tok@fm}{\def\PY@tc##1{\textcolor[rgb]{0.02,0.16,0.49}{##1}}}
\@namedef{PY@tok@vc}{\def\PY@tc##1{\textcolor[rgb]{0.73,0.38,0.84}{##1}}}
\@namedef{PY@tok@vg}{\def\PY@tc##1{\textcolor[rgb]{0.73,0.38,0.84}{##1}}}
\@namedef{PY@tok@vi}{\def\PY@tc##1{\textcolor[rgb]{0.73,0.38,0.84}{##1}}}
\@namedef{PY@tok@vm}{\def\PY@tc##1{\textcolor[rgb]{0.73,0.38,0.84}{##1}}}
\@namedef{PY@tok@sa}{\def\PY@tc##1{\textcolor[rgb]{0.25,0.44,0.63}{##1}}}
\@namedef{PY@tok@sb}{\def\PY@tc##1{\textcolor[rgb]{0.25,0.44,0.63}{##1}}}
\@namedef{PY@tok@sc}{\def\PY@tc##1{\textcolor[rgb]{0.25,0.44,0.63}{##1}}}
\@namedef{PY@tok@dl}{\def\PY@tc##1{\textcolor[rgb]{0.25,0.44,0.63}{##1}}}
\@namedef{PY@tok@s2}{\def\PY@tc##1{\textcolor[rgb]{0.25,0.44,0.63}{##1}}}
\@namedef{PY@tok@sh}{\def\PY@tc##1{\textcolor[rgb]{0.25,0.44,0.63}{##1}}}
\@namedef{PY@tok@s1}{\def\PY@tc##1{\textcolor[rgb]{0.25,0.44,0.63}{##1}}}
\@namedef{PY@tok@mb}{\def\PY@tc##1{\textcolor[rgb]{0.13,0.50,0.31}{##1}}}
\@namedef{PY@tok@mf}{\def\PY@tc##1{\textcolor[rgb]{0.13,0.50,0.31}{##1}}}
\@namedef{PY@tok@mh}{\def\PY@tc##1{\textcolor[rgb]{0.13,0.50,0.31}{##1}}}
\@namedef{PY@tok@mi}{\def\PY@tc##1{\textcolor[rgb]{0.13,0.50,0.31}{##1}}}
\@namedef{PY@tok@il}{\def\PY@tc##1{\textcolor[rgb]{0.13,0.50,0.31}{##1}}}
\@namedef{PY@tok@mo}{\def\PY@tc##1{\textcolor[rgb]{0.13,0.50,0.31}{##1}}}
\@namedef{PY@tok@ch}{\let\PY@it=\textit\def\PY@tc##1{\textcolor[rgb]{0.25,0.50,0.56}{##1}}}
\@namedef{PY@tok@cm}{\let\PY@it=\textit\def\PY@tc##1{\textcolor[rgb]{0.25,0.50,0.56}{##1}}}
\@namedef{PY@tok@cpf}{\let\PY@it=\textit\def\PY@tc##1{\textcolor[rgb]{0.25,0.50,0.56}{##1}}}
\@namedef{PY@tok@c1}{\let\PY@it=\textit\def\PY@tc##1{\textcolor[rgb]{0.25,0.50,0.56}{##1}}}

\def\PYZus{\char`\_}
\def\PYZob{\char`\{}
\def\PYZcb{\char`\}}

\def\PYZsh{\char`\#}

\def\PYZhy{\char`\-}

\def\PYZdq{\char`\"}

\makeatother

\providecommand*{\DUfootnotemark}[3]{%
  \raisebox{1em}{\hypertarget{#1}{}}%
  \hyperlink{#2}{\textsuperscript{#3}}%
}
\providecommand{\DUfootnotetext}[4]{%
  \begingroup%
  \renewcommand{\thefootnote}{%
    \protect\raisebox{1em}{\protect\hypertarget{#1}{}}%
    \protect\hyperlink{#2}{#3}}%
  \footnotetext{#4}%
  \endgroup%
}

\providecommand*{\DUrole}[2]{%
  \ifcsname docutilsrole#1\endcsname%
    \csname docutilsrole#1\endcsname{#2}%
  \else
    \csname DUrole#1\endcsname{#2}%
  \fi%
}

\providecommand*{\DUroletitlereference}[1]{\textsl{#1}}

\ifthenelse{\isundefined{\hypersetup}}{
  \usepackage[colorlinks=true,linkcolor=blue,urlcolor=blue]{hyperref}
  \usepackage{bookmark}
  \urlstyle{same} 
}{}

\begin{document}
\newcounter{footnotecounter}\title{atoMEC: An open-source average-atom Python code}\author{Timothy J. Callow$^{\setcounter{footnotecounter}{3}\fnsymbol{footnotecounter}\setcounter{footnotecounter}{4}\fnsymbol{footnotecounter}\setcounter{footnotecounter}{1}\fnsymbol{footnotecounter}}$%
          \setcounter{footnotecounter}{1}\thanks{\fnsymbol{footnotecounter} %
          Corresponding author: \protect\href{mailto:t.callow@hzdr.de}{t.callow@hzdr.de}}\setcounter{footnotecounter}{3}\thanks{\fnsymbol{footnotecounter} Center for Advanced Systems Understanding (CASUS), D-02826 Görlitz, Germany}\setcounter{footnotecounter}{4}\thanks{\fnsymbol{footnotecounter} Helmholtz-Zentrum Dresden-Rossendorf, D-01328 Dresden, Germany}, Daniel Kotik$^{\setcounter{footnotecounter}{3}\fnsymbol{footnotecounter}\setcounter{footnotecounter}{4}\fnsymbol{footnotecounter}}$, Eli Kraisler$^{\setcounter{footnotecounter}{5}\fnsymbol{footnotecounter}}$\setcounter{footnotecounter}{5}\thanks{\fnsymbol{footnotecounter} Fritz Haber Center for Molecular Dynamics and Institute of Chemistry, The Hebrew University of Jerusalem, 9091401 Jerusalem, Israel}, Attila Cangi$^{\setcounter{footnotecounter}{3}\fnsymbol{footnotecounter}\setcounter{footnotecounter}{4}\fnsymbol{footnotecounter}}$\thanks{%

          \noindent%
          Copyright\,\copyright\,2022 Timothy J. Callow et al. This is an open-access article distributed under the terms of the Creative Commons Attribution License, which permits unrestricted use, distribution, and reproduction in any medium, provided the original author and source are credited.%
        }}\maketitle
          \renewcommand{\leftmark}{PROC. OF THE 21st PYTHON IN SCIENCE CONF. (SCIPY 2022)}
          \renewcommand{\rightmark}{ATOMEC: AN OPEN-SOURCE AVERAGE-ATOM PYTHON CODE}
        
\InputIfFileExists{page_numbers.tex}{}{}
\newcommand*{\docutilsroleref}{\ref}
\newcommand*{\docutilsrolelabel}{\label}
\newcommand*\DUrolecode[1]{#1}
\providecommand*\DUrolecite[1]{\cite{#1}}

\begin{abstract}Average-atom models are an important tool in studying matter under extreme conditions, such as those conditions experienced in planetary cores, brown and white dwarfs, and during inertial confinement fusion.
In the right context, average-atom models can yield results with similar accuracy to simulations which require orders of magnitude more computing time, and thus can greatly reduce financial and environmental costs.
Unfortunately, due to the wide range of possible models and approximations, and the lack of open-source codes, average-atom models can at times appear inaccessible.
In this paper, we present our open-source average-atom code, \href{https://github.com/atomec-project/atoMEC}{atoMEC}.
We explain the aims and structure of atoMEC to illuminate the different stages and options in an average-atom calculation, and to facilitate community contributions.
We also discuss the use of various open-source Python packages in atoMEC, which have expedited its development.\end{abstract}

\subsection{Introduction%
  \label{introduction}%
}

The study of matter under extreme conditions — materials exposed to high temperatures, high pressures, or strong electromagnetic fields — is critical to our understanding of many important scientific and technological processes, such as nuclear fusion and various astrophysical and planetary physics phenomena \DUrole{cite}{MEC_linac}.
Of particular interest within this broad field is the warm dense matter (WDM) regime, which is typically characterized by temperatures in the range of $10^3 -10^6$ degrees (Kelvin), and densities ranging from dense gases to highly compressed solids ($\sim 0.01 - 1000\ \textrm{g cm}^{-3}$) \DUrole{cite}{Bonitz_20}.
In this regime, it is important to account for the quantum mechanical nature of the electrons (and in some cases, also the nuclei). Therefore conventional methods from plasma physics, which either neglect quantum effects or treat them coarsely, are usually not sufficiently accurate.
On the other hand, methods from condensed-matter physics and quantum chemistry, which account fully for quantum interactions, typically target the ground-state only, and become computationally intractable for systems at high temperatures.

Nevertheless, there are methods which can, in principle, be applied to study materials at any given temperature and density whilst formally accounting for quantum interactions. These methods are often denoted as \textquotedbl{}first-principles\textquotedbl{} because, formally speaking, they yield the exact properties of the system, under certain well-founded theoretical approximations.
Density-functional theory (DFT), initially developed as a ground-state theory \DUrole{cite}{HK64,KS65} but later extended to non-zero temperatures \DUrole{cite}{M65,FT_DFT_exact}, is one such theory and has been used extensively to study materials under WDM conditions \DUrole{cite}{graziani_14}.
Even though DFT reformulates the Schrödinger equation in a computationally efficient manner \DUrole{cite}{Kohn_Nobel_lecture}, the cost of running calculations becomes prohibitively expensive at higher temperatures. Formally, it scales as $\mathcal{O}(N^3 \tau^3)$, with $N$ the particle number (which usually also increases with temperature) and $\tau$ the temperature \DUrole{cite}{stoc_DFT}.
This poses a serious computational challenge in the WDM regime.
Furthermore, although DFT is a formally exact theory, in practice it relies on approximations for the so-called \textquotedbl{}exchange-correlation\textquotedbl{} energy, which is, roughly speaking, responsible for simulating all the quantum interactions between electrons.
Existing exchange-correlation approximations have not been rigorously tested under WDM conditions.
An alternative method used in the WDM community is path-integral Monte–Carlo \DUrole{cite}{DGB18}, which yields essentially exact properties; however, it is even more limited by computational cost than DFT, and becomes unfeasibly expensive at lower temperatures due to the fermion sign problem.

It is therefore of great interest to reduce the computational complexity of the aforementioned methods.
The use of graphics processing units in DFT calculations is becomingly increasingly common, and has been shown to offer significant speed-ups relative to conventional calculations using central processing units \DUrole{cite}{GPUs_1,GPUs_2}.
Some other examples of promising developments to reduce the cost of DFT calculations include machine-learning-based solutions \DUrole{cite}{ML_DFT_1,ML_DFT_2,mala} and stochastic DFT \DUrole{cite}{stoc_DFT,stoc_DFT_2}.
However, in this paper, we focus on an alternative class of models known as \textquotedbl{}average-atom\textquotedbl{} models.
Average-atom models have a long history in plasma physics \DUrole{cite}{PRR_AA}: they account for quantum effects, typically using DFT, but reduce the complex system of interacting electrons and nuclei to a single atom immersed in a plasma (the \textquotedbl{}average\textquotedbl{} atom).
An illustration of this principle (reduced to two dimensions for visual purposes) is shown in Fig. 1.
This significantly reduces the cost relative to a full DFT simulation, because the particle number is restricted to the number of electrons per nucleus, and spherical symmetry is exploited to reduce the three-dimensional problem to one dimension.

Naturally, to reduce the complexity of the problem as described, various approximations must be introduced.
It is important to understand these approximations and their limitations for average-atom models to have genuine predictive capabilities.
Unfortunately, this is not always the case: although average-atom models share common concepts, there is no unique formal theory underpinning them.
Therefore a variety of models and codes exist, and it is not typically clear which models can be expected to perform most accurately under which conditions.
In a previous paper \DUrole{cite}{PRR_AA}, we addressed this issue by deriving an average-atom model from first principles, and comparing the impact of different approximations within this model on some common properties.

In this paper, we focus on computational aspects of average-atom models for WDM.
We introduce atoMEC \DUrole{cite}{atoMEC_zenodo}: an open-source average-\textbf{ato}m code for studying \textbf{M}atter under \textbf{E}xtreme \textbf{C}onditions.
One of the main aims of atoMEC is to improve the accessibility and understanding of average-atom models.
To the best of our knowledge, open-source average-atom codes are in scarce supply: with atoMEC, we aim to provide a tool that people can use to run average-atom simulations and also to add their own models, which should facilitate comparisons of different approximations.
The relative simplicity of average-atom codes means that they are not only efficient to run, but also efficient to develop: this means, for example, that they can be used as a test-bed for new ideas that could be later implemented in full DFT codes, and are also accessible to those without extensive prior expertise, such as students.
atoMEC aims to facilitate development by following good practice in software engineering (for example extensive documentation), a careful design structure, and of course through the choice of Python and its widely used scientific stack, in particular the NumPy \DUrole{cite}{numpy} and SciPy \DUrole{cite}{scipy} libraries.

This paper is structured as follows: in the next section, we briefly review the key theoretical points which are important to understand the functionality of atoMEC, assuming no prior physical knowledge of the reader.
Following that, we present the key functionality of atoMEC, discuss the code structure and algorithms, and explain how these relate to the theoretical aspects introduced.
Finally, we present an example case study: we consider helium under the conditions often experienced in the outer layers of a white dwarf star, and probe the behavior of a few important properties, namely the band-gap, pressure, and ionization degree.\begin{figure}[]\noindent\makebox[\columnwidth][c]{\includegraphics[scale=1.00]{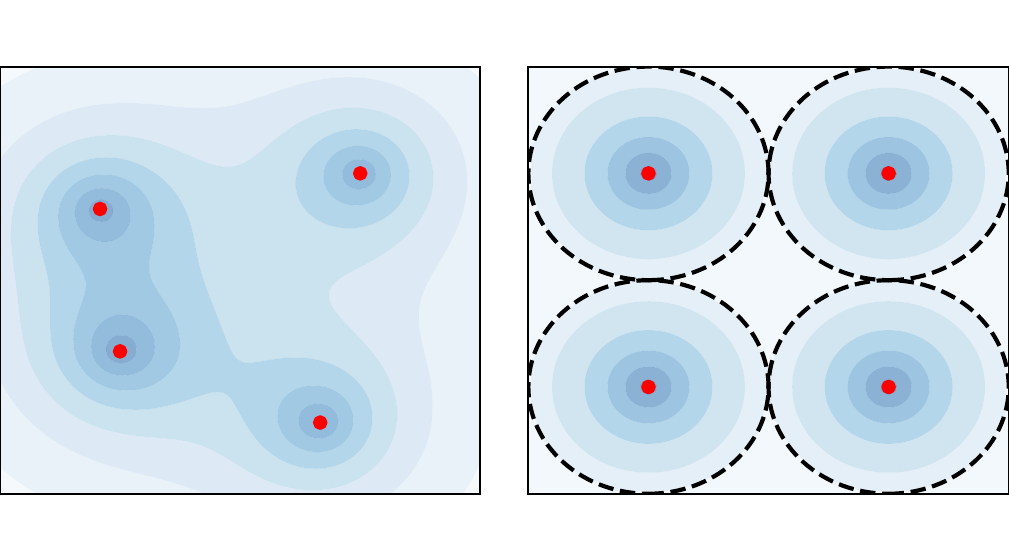}}
\caption{Illustration of the average-atom concept. The many-body and fully-interacting system of electron density (shaded blue) and nuclei (red points) on the left is mapped into the much simpler system of independent atoms on the right.
Any of these identical atoms represents the \textquotedbl{}average-atom\textquotedbl{}. The effects of interaction from neighboring atoms are implicitly accounted for in an approximate manner through the choice of boundary conditions.}
\end{figure}

\subsection{Theoretical background%
  \label{theoretical-background}%
}

Properties of interest in the warm dense matter regime include the equation-of-state data, which is the relation between the density, energy, temperature and pressure of a material \DUrole{cite}{hugoniot}; the mean ionization state and the electron ionization energies, which tell us about how tightly bound the electrons are to the nuclei; and the electrical and thermal conductivities.
These properties yield information pertinent to our understanding of stellar and planetary physics, the Earth's core, inertial confinement fusion, and more besides.
To exactly obtain these properties, one needs (in theory) to determine the thermodynamic ensemble of the quantum states (the so-called \emph{wave-functions}) representing the electrons and nuclei.
Fortunately, they can be obtained with reasonable accuracy using models such as average-atom models; in this section, we elaborate on how this is done.

We shall briefly review the key theory underpinning the type of average-atom model implemented in atoMEC. This is intended for readers without a background in quantum mechanics, to give some context to the purposes and mechanisms of the code.
For a comprehensive derivation of this average-atom model, we direct readers to Ref. \DUrole{cite}{PRR_AA}.
The average-atom model we shall describe falls into a class of models known as \emph{ion-sphere} models, which are the simplest (and still most widely used) class of average-atom model.
There are alternative (more advanced) classes of model such as \emph{ion-correlation} \DUrole{cite}{ioncorrelation} and \emph{neutral pseudo-atom} models \DUrole{cite}{NPA} which we have not yet implemented in atoMEC, and thus we do not elaborate on them here.

As demonstrated in Fig. 1, the idea of the ion-sphere model is to map a fully-interacting system of many electrons and nuclei into a set of independent atoms which do not interact explicitly with any of the other spheres.
Naturally, this depends on several assumptions and approximations, but there is formal justification for such a mapping \DUrole{cite}{PRR_AA}.
Furthermore, there are many examples in which average-atom models have shown good agreement with more accurate simulations and experimental data \DUrole{cite}{AA_pressure}, which further justifies this mapping.

Although the average-atom picture is significantly simplified relative to the full many-body problem, even determining the wave-functions and their ensemble weights for an atom at finite temperature is a complex problem.
Fortunately, DFT reduces this complexity further, by establishing that the electron \emph{density} — a far less complex entity than the wave-functions — is sufficient to determine all physical observables.
The most popular formulation of DFT, known as Kohn–Sham DFT (KS-DFT) \DUrole{cite}{KS65}, allows us to construct the \emph{fully-interacting} density from a \emph{non-interacting} system of electrons, simplifying the problem further still.
Due to the spherical symmetry of the atom, the non-interacting electrons — known as KS electrons (or KS orbitals) — can be represented as a wave-function that is a product of radial and angular components,\begin{equation}
\label{eq:phi}
\phi_{nlm}(\mathbf{r}) = X_{nl}(r) Y_l^m(\theta, \phi)\,,
\end{equation}where $n,\ l,\ \textrm{and}\ m$ are the \emph{quantum numbers} of the orbitals, which come from the fact that the wave-function is an eigenfunction of the Hamiltonian operator, and $Y_l^m(\theta, \phi)$ are the spherical harmonic functions.\DUfootnotemark{id1}{f2}{1} The radial coordinate $r$ represents the absolute distance from the nucleus.%
\DUfootnotetext{f2}{id1}{1}{\phantomsection\label{f2}Please note that the notation in Eq. (\DUrole{ref}{eq:phi}) does not imply Einstein summation notation. All summations in this paper are written explicitly; Einstein summation notation is not used.}

We therefore only need to determine the radial KS orbitals $X_{nl}(r)$.
These are determined by solving the radial KS equation, which is similar to the Schrödinger equation for a non-interacting system, with an additional term in the potential to mimic the effects of electron-electron interaction (within the single atom).
The radial KS equation is given by:\begin{equation}
\label{eq:kseqn}
\left[- \left(\frac{\textrm{d}^2}{\textrm{d}r^2} + \frac{2}{r}\frac{\textrm{d}}{\textrm{d}r} - \frac{l(l+1)}{r^2} \right) + v_\textrm{s}[n](r) \right] X_{nl}(r) = \epsilon_{nl} X_{nl}(r).
\end{equation}We have written the above equation in a way that emphasizes that it is an eigenvalue equation, with the eigenvalues $\epsilon_{nl}$ being the energies of the KS orbitals.

On the left-hand side, the terms in the round brackets come from the kinetic energy operator acting on the orbitals.
The $v_\textrm{s}[n](r)$ term is the KS potential, which itself is composed of three different terms,\begin{equation}
\label{eq:kspot}
v_{\textrm{s}}[n](r) = -\frac{Z}{r} + 4\pi \int_0^{R_\textrm{WS}} \textrm{d}{x} \frac{n(x)x^2}{\max(r,x)} + \frac{\delta F_\textrm{xc}[n]}{\delta n(r)}\,,
\end{equation}where $R_\textrm{WS}$ is the radius of the atomic sphere, $n(r)$ is the electron density, $Z$ the nuclear charge, and $F_\textrm{xc}[n]$ the exchange-correlation free energy functional.
Thus the three terms in the potential are respectively the electron-nuclear attraction, the classical Hartree repulsion, and the exchange-correlation (xc) potential.

We note that the KS potential and its constituents are functionals of the electron density $n(r)$.
Were it not for this dependence on the density, solving Eq. \DUrole{ref}{eq:kseqn} just amounts to solving an ordinary linear differential equation (ODE).
However, the electron density is in fact constructed from the orbitals in the following way,\begin{equation}
\label{eq:dens}
n(r) = 2\sum_{nl}(2l+1) f_{nl}(\epsilon_{nl},\mu,\tau) |X_{nl}(r)|^2\,,
\end{equation}where $f_{nl}(\epsilon_{nl},\mu,\tau)$ is the Fermi–Dirac distribution, given by\begin{equation}
\label{eq:fdocc}
f_{nl}(\epsilon_{nl},\mu,\tau) = \frac{1}{1+e^{(\epsilon_{nl}-\mu)/\tau}}\,,
\end{equation}where $\tau$ is the temperature, and $\mu$ is the chemical potential, which is determined by fixing the number of electrons to be equal to a pre-determined value $N_\textrm{e}$ (typically equal to the nuclear charge $Z$).
The Fermi–Dirac distribution therefore assigns weights to the KS orbitals in the construction of the density, with the weight depending on their energy.

Therefore, the KS potential that determines the KS orbitals via the ODE (\DUrole{ref}{eq:kseqn}), is itself dependent on the KS orbitals.
Consequently, the KS orbitals and their dependent quantities (the density and KS potential) must be determined via a so-called self-consistent field (SCF) procedure.
An initial guess for the orbitals, $X_{nl}^0(r)$, is used to construct the initial density $n^0(r)$ and potential $v_\textrm{s}^0(r)$.
The ODE (\DUrole{ref}{eq:kseqn}) is then solved to update the orbitals.
This process is iterated until some appropriately chosen quantities — in atoMEC the total free energy, density and KS potential — are converged, i.e. $n^{i+1}(r)=n^i(r),\ v_\textrm{s}^{i+1}(r)=v_\textrm{s}^i(r),\ F^{i+1} = F^i$, within some reasonable numerical tolerance.
In Fig. 2, we illustrate the life-cycle of the average-atom model described so far, including the SCF procedure.
On the left-hand side of this figure, we show the physical choices and mathematical operations, and on the right-hand side, the representative classes and functions in atoMEC.
In the following section, we shall discuss some aspects of this figure in more detail.

Some quantities obtained from the completion of the SCF procedure are directly of interest.
For example, the energy eigenvalues $\epsilon_{nl}$ are related to the electron ionization energies, i.e. the amount of energy required to excite an electron bound to the nucleus to being a free (conducting) electron.
These predicted ionization energies can be used, for example, to help understand ionization potential depression, an important but somewhat controversial effect in WDM \DUrole{cite}{IPDdepression}.
Another property that can be straightforwardly obtained from the energy levels and their occupation numbers is the mean ionization state $\bar{Z}$\DUfootnotemark{id2}{f1}{2},\begin{equation}
\label{eq:MIS}
\bar{Z} = \sum_{n,l} (2l+1) f_{nl}(\epsilon_{nl}, \mu, \tau)
\end{equation}which is an important input parameter for various models, such as adiabats which are used to model inertial confinement fusion \DUrole{cite}{ICFadiabats}.%
\DUfootnotetext{f1}{id2}{2}{\phantomsection\label{f1}The summation in Eq. (\DUrole{ref}{eq:MIS}) is often shown as an integral because the energies above a certain threshold form a continuous distribution (in most models).}

Various other interesting properties can also be calculated following some post-processing of the output of an SCF calculation, for example the pressure exerted by the electrons and ions.
Furthermore, response properties, i.e. those resulting from an external perturbation like a laser pulse, can also be obtained from the output of an SCF cycle.
These properties include, for example, electrical conductivities \DUrole{cite}{AA_KG} and dynamical structure factors \DUrole{cite}{AA_DSF}.\begin{figure*}[]\noindent\makebox[\textwidth][c]{\includegraphics[scale=0.90]{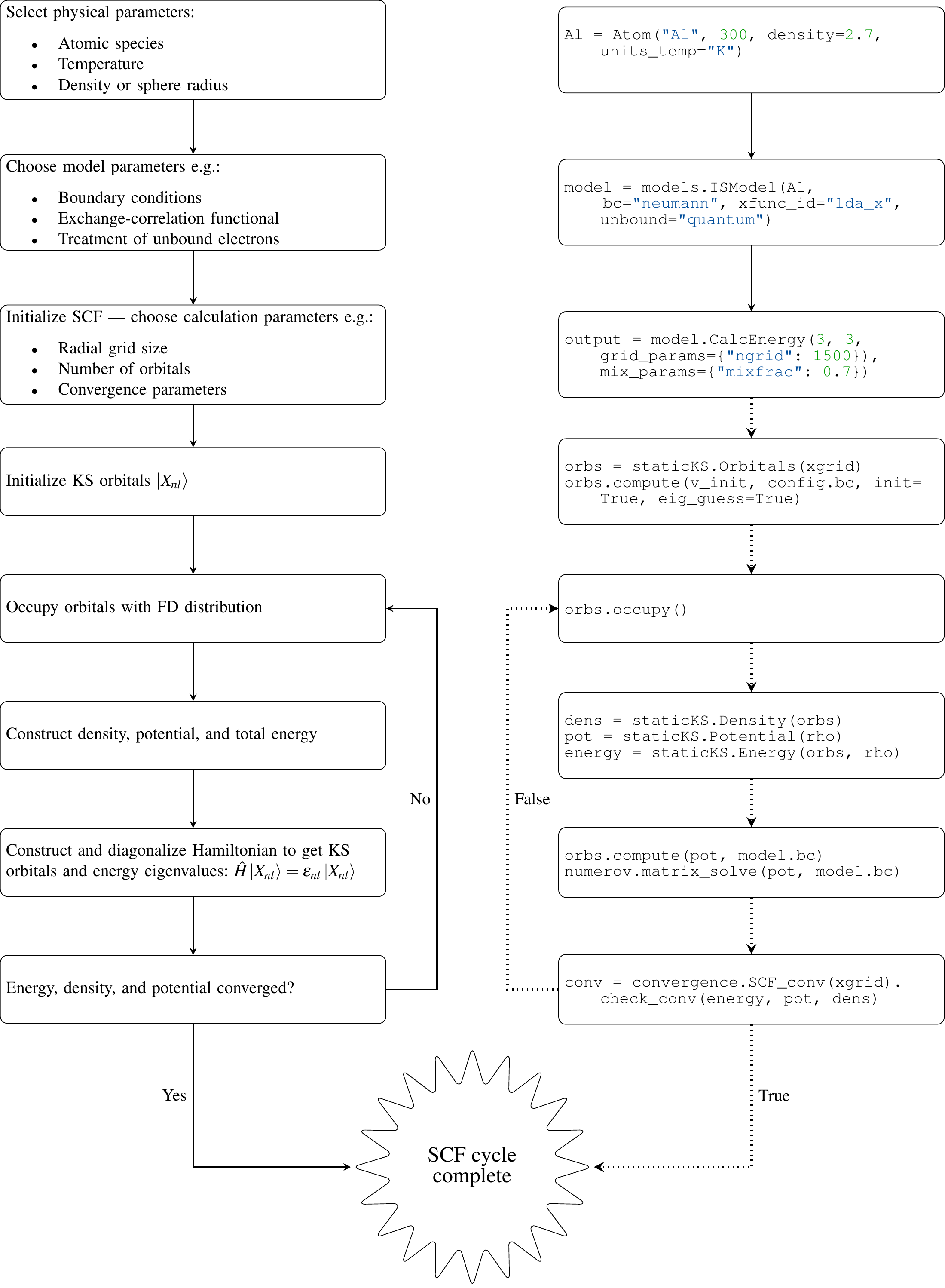}}
\caption{Schematic of the average-atom model set-up and the self-consistent field (SCF) cycle.
On the left-hand side, the physical choices and mathematical operations that define the model and SCF cycle are shown.
On the right-hand side, the (higher-order) functions and classes in atoMEC corresponding to the items on the left-hand side are shown.
Some liberties are taken with the code snippets in the right-hand column of the figure to improve readability; more precisely, some non-crucial intermediate steps are not shown, and some parameters are also not shown or simplified.
The dotted lines represent operations that are taken care of within the \texttt{\DUrole{code}{models.CalcEnergy}} function, but are shown nevertheless to improve understanding.}
\end{figure*}

\subsection{Code structure and details%
  \label{code-structure-and-details}%
}

In the following sections, we describe the structure of the code in relation to the physical problem being modeled.
Average-atom models typically rely on various parameters and approximations.
In atoMEC, we have tried to structure the code in a way that makes clear which parameters come from the physical problem studied compared to choices of the model and numerical or algorithmic choices.

\subsubsection{\DUroletitlereference{atoMEC.Atom}: Physical parameters%
  \label{atomec-atom-physical-parameters}%
}

The first step of any simulation in WDM (which also applies to simulations in science more generally) is to define the physical parameters of the problem.
These parameters are unique in the sense that, if we had an exact method to simulate the real system, then for each combination of these parameters there would be a unique solution.
In other words, regardless of the model — be it average atom or a different technique — these parameters are always required and are independent of the model.

In average-atom models, there are typically three parameters defining the physical problem, which are:
\begin{itemize}
\item 

the \textbf{atomic species};
\item 

the \textbf{temperature} of the material, $\tau$;
\item 

the \textbf{mass density} of the material, $\rho_\textrm{m}$.\end{itemize}

The mass density also directly corresponds to the mean distance between two nuclei (atomic centers), which in the average-atom model is equal to twice the radius of the atomic sphere, $R_\textrm{WS}$.
An additional physical parameter not mentioned above is the \textbf{net charge} of the material being considered, i.e. the difference between the nuclear charge $Z$ and the electron number $N_\textrm{e}$.
However, we usually assume zero net charge in average-atom simulations (i.e. the number of electrons is equal to the atomic charge).

In atoMEC, these physical parameters are controlled by the \texttt{\DUrole{code}{Atom}} object.
As an example, we consider aluminum under ambient conditions, i.e. at room temperature, $\tau=300\ \textrm{K}$, and normal metallic density, $\rho_\textrm{m}=2.7\ \textrm{g cm}^{-3}$.
We set this up as:\vspace{1mm}
\begin{Verbatim}[commandchars=\\\{\},fontsize=\footnotesize]
\PY{k+kn}{from} \PY{n+nn}{atoMEC} \PY{k+kn}{import} \PY{n}{Atom}
\PY{n}{Al} \PY{o}{=} \PY{n}{Atom}\PY{p}{(}\PY{l+s+s2}{\PYZdq{}}\PY{l+s+s2}{Al}\PY{l+s+s2}{\PYZdq{}}\PY{p}{,} \PY{l+m+mi}{300}\PY{p}{,} \PY{n}{density}\PY{o}{=}\PY{l+m+mf}{2.7}\PY{p}{,} \PY{n}{units\PYZus{}temp}\PY{o}{=}\PY{l+s+s2}{\PYZdq{}}\PY{l+s+s2}{K}\PY{l+s+s2}{\PYZdq{}}\PY{p}{)}
\end{Verbatim}
\vspace{1mm}
\begin{figure}[]\noindent\makebox[\columnwidth][c]{\includegraphics[width=\columnwidth]{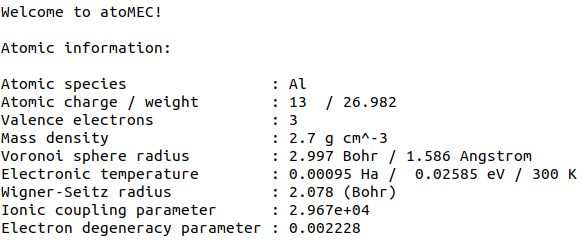}}
\caption{Auto-generated print statement from calling the \texttt{\DUrole{code}{atoMEC.Atom}} object.}
\end{figure}By default, the above code automatically prints the output seen in Fig. 3. We see that the first two arguments of the \texttt{\DUrole{code}{Atom}} object are the chemical symbol of the element being studied, and the temperature.
In addition, at least one of \textquotedbl{}density\textquotedbl{} or \textquotedbl{}radius\textquotedbl{} must be specified.
In atoMEC, the default (and only permitted) units for the mass density are $\textrm{g cm}^{-3}$; \emph{all} other input and output units in atoMEC are by default Hartree atomic units, and hence we specify \textquotedbl{}K\textquotedbl{} for Kelvin.

The information in Fig. 3 displays the chosen parameters in units commonly used in the plasma and condensed-matter physics communities, as well as some other information directly obtained from these parameters.
The chemical symbol (\textquotedbl{}Al\textquotedbl{} in this case) is passed to the mendeleev library \DUrole{cite}{mendeleev2014} to generate this data, which is used later in the calculation.

This initial stage of the average-atom calculation, i.e. the specification of physical parameters and initialization of the \texttt{\DUrole{code}{Atom}} object, is shown in the top row at the top of Fig. 2.

\subsubsection{\DUroletitlereference{atoMEC.models}: Model parameters%
  \label{atomec-models-model-parameters}%
}

After the physical parameters are set, the next stage of the average-atom calculation is to choose the model and approximations within that class of model.
As discussed, so far the only class of model implemented in atoMEC is the ion-sphere model.
Within this model, there are still various choices to be made by the user.
In some cases, these choices make little difference to the results, but in other cases they have significant impact.
The user might have some physical intuition as to which is most important, or alternatively may want to run the same physical parameters with several different model parameters to examine the effects.
Some choices available in atoMEC, listed approximately in decreasing order of impact (but this can depend strongly on the system under consideration), are:
\begin{itemize}
\item 

the \textbf{boundary conditions} used to solve the KS equations;
\item 

the treatment of the \textbf{unbound electrons}, which means those electrons not tightly bound to the nucleus, but rather delocalized over the whole atomic sphere;
\item 

the choice of \textbf{exchange} and \textbf{correlation} functionals, the central approximations of DFT \DUrole{cite}{xc_review};
\item 

the \textbf{spin} polarization and magnetization.\end{itemize}

We do not discuss the theory and impact of these different choices in this paper. Rather, we direct readers to Refs. \DUrole{cite}{PRR_AA} and \DUrole{cite}{arxiv_KG} in which all of these choices are discussed.

In atoMEC, the ion-sphere model is controlled by the \texttt{\DUrole{code}{models.ISModel}} object. Continuing with our aluminum example, we choose the so-called \textquotedbl{}neumann\textquotedbl{} boundary condition, with a \textquotedbl{}quantum\textquotedbl{} treatment of the unbound electrons, and choose the LDA exchange functional (which is also the default). This model is set up as:\vspace{1mm}
\begin{Verbatim}[commandchars=\\\{\},fontsize=\footnotesize]
\PY{k+kn}{from} \PY{n+nn}{atoMEC} \PY{k+kn}{import} \PY{n}{models}
\PY{n}{model} \PY{o}{=} \PY{n}{models}\PY{o}{.}\PY{n}{ISModel}\PY{p}{(}\PY{n}{Al}\PY{p}{,} \PY{n}{bc}\PY{o}{=}\PY{l+s+s2}{\PYZdq{}}\PY{l+s+s2}{neumann}\PY{l+s+s2}{\PYZdq{}}\PY{p}{,}
             \PY{n}{xfunc\PYZus{}id}\PY{o}{=}\PY{l+s+s2}{\PYZdq{}}\PY{l+s+s2}{lda\PYZus{}x}\PY{l+s+s2}{\PYZdq{}}\PY{p}{,} \PY{n}{unbound}\PY{o}{=}\PY{l+s+s2}{\PYZdq{}}\PY{l+s+s2}{quantum}\PY{l+s+s2}{\PYZdq{}}\PY{p}{)}
\end{Verbatim}
\vspace{1mm}
By default, the above code prints the output shown in Fig. 4. The first (and only mandatory) input parameter to the \texttt{\DUrole{code}{models.ISModel}} object is the \texttt{\DUrole{code}{Atom}} object that we generated earlier.
Together with the optional \texttt{\DUrole{code}{spinpol}} and \texttt{\DUrole{code}{spinmag}} parameters in the \texttt{\DUrole{code}{models.ISModel}} object, this sets either the total number of electrons (\texttt{\DUrole{code}{spinpol=False}}) or the number of electrons in each spin channel (\texttt{\DUrole{code}{spinpol=True}}).

The remaining information displayed in Fig. 4 shows directly the chosen model parameters, or the default values where these parameters are not specified.
The exchange and correlation functionals - set by the parameters \texttt{\DUrole{code}{xfunc\_id}} and \texttt{\DUrole{code}{cfunc\_id}} - are passed to the LIBXC library \DUrole{cite}{libxc_2018} for processing.
So far, only the \textquotedbl{}local density\textquotedbl{} family of approximations is available in atoMEC, and thus the default values are usually a sensible choice.
For more information on exchange and correlation functionals, there are many reviews in the literature, for example Ref. \DUrole{cite}{xc_review}.

This stage of the average-atom calculation, i.e. the specification of the model and the choices of approximation within that, is shown in the second row of Fig. 2.\begin{figure}[]\noindent\makebox[\columnwidth][l]{\includegraphics[scale=0.45]{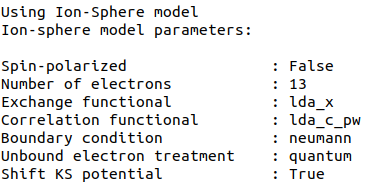}}
\caption{Auto-generated print statement from calling the \texttt{\DUrole{code}{models.ISModel}} object.}
\end{figure}

\subsubsection{\DUroletitlereference{ISModel.CalcEnergy}: SCF calculation and numerical parameters%
  \label{ismodel-calcenergy-scf-calculation-and-numerical-parameters}%
}

Once the physical parameters and model have been defined, the next stage in the average-atom calculation (or indeed any DFT calculation) is the SCF procedure.
In atoMEC, this is invoked by the \texttt{\DUrole{code}{ISModel.CalcEnergy}} function.
This function is called \texttt{\DUrole{code}{CalcEnergy}} because it finds the KS orbitals (and associated KS density) which minimize the total free energy.

Clearly, there are various mathematical and algorithmic choices in this calculation.
These include, for example: the basis in which the KS orbitals and potential are represented, the algorithm used to solve the KS equations (\DUrole{ref}{eq:kseqn}), and how to ensure smooth convergence of the SCF cycle.
In atoMEC, the SCF procedure currently follows a single pre-determined algorithm, which we briefly review below.

In atoMEC, we represent the radial KS quantities (orbitals, density and potential) on a logarithmic grid, i.e. $x=\log(r)$.
Furthermore, we make a transformation of the orbitals $P_{nl}(x) = X_{nl}(x)e^{x/2}$. Then the equations to be solved become:\begin{eqnarray}
\label{eq:logkseqn}
\frac{\textrm{d}^2 P_{nl}(x)}{\textrm{d}x^2} - 2e^{2x}(W(x)-\epsilon_{nl})P_{nl}(x)=0 \\
W(x) = v_\textrm{s}[n](x) + \frac{1}{2}\left(l+\frac{1}{2}\right)^2 e^{-2x}\,.
\end{eqnarray}In atoMEC, we solve the KS equations using a matrix implementation of Numerov's algorithm \DUrole{cite}{matrix_numerov}.
This means we diagonalize the following equation:\begin{eqnarray}
\label{eq:ham}
\hat{H}\vec{P} &&= \vec{\epsilon} \hat{B} \vec{P}\,,\ \textrm{where} \\
\hat{H} &&= \hat{T} + \hat{B} + W_\textrm{s}(\vec{x})\,, \\
\hat{T} &&= -\frac{1}{2} e^{-2\vec{x}} \hat{A}\,, \\
\hat{A} &&= \frac{\hat{I}_{-1} -2\hat{I}_0 + \hat{I}_1}{\textrm{d}x^2}\,,\textrm{and} \\
\hat{B} &&= \frac{\hat{I}_{-1} +10\hat{I}_0 + \hat{I}_1}{12}\,,
\end{eqnarray}In the above, $\hat{I}_{-1/0/1}$ are lower shift, identify, and upper shift matrices.

The Hamiltonian matrix $\hat{H}$ is sparse and we only seek a subset of eigenstates with lower energies: therefore there is no need to perform a full diagonalization, which scales as $\mathcal{O}(N^3)$, with $N$ being the size of the radial grid.
Instead, we use SciPy's sparse matrix diagonalization function \texttt{\DUrole{code}{scipy.sparse.linalg.eigs}}, which scales more efficiently and allows us to go to larger grid sizes.

After each step in the SCF cycle, the relative changes in the free energy $F$, density $n(r)$ and potential $v_\textrm{s}(r)$ are computed.
Specifically, the quantities computed are\begin{eqnarray}
\label{eq:conv}
\Delta F &&= \left|\frac{F^{i}-F^{i-1}}{F^{i}}\right| \\
 \Delta n &&= \frac{\int \mathrm{d}r|n^i(r)-n^{i-1}(r)|}{\int \mathrm{d}r n^i(r)}\\
 \Delta v &&= \frac{\int \mathrm{d}r|v^i_\textrm{s}(r)-v_\textrm{s}^{i-1}(r)|}{\int \mathrm{d}r v_\textrm{s}^i(r)}\,.
\end{eqnarray}Once all three of these metrics fall below a certain threshold, the SCF cycle is considered converged and the calculation finishes.

The SCF cycle is an example of a non-linear system and thus is prone to chaotic (non-convergent) behavior.
Consequently a range of techniques have been developed to ensure convergence \DUrole{cite}{SCFconvergence}.
Fortunately, the tendency for calculations not to converge becomes less likely for temperatures above zero (and especially as temperatures increase).
Therefore we have implemented only a simple linear mixing scheme in atoMEC.
The potential used in each diagonalization step of the SCF cycle is not simply the one generated from the most recent density, but a mix of that potential and the previous one,\begin{equation}
\label{eq:potmix}
v_\textrm{s}^{(i)}(r) = \alpha v_\textrm{s}^{i}(r) + (1 - \alpha) v_\textrm{s}^{i-1}(r)\,.
\end{equation}In general, a lower value of the mixing fraction $\alpha$ makes the SCF cycle more stable, but requires more iterations to converge.
Typically a choice of $\alpha\approx 0.5$ gives a reasonable balance between speed and stability.

We can thus summarize the key parameters in an SCF calculation as follows:
\begin{itemize}
\item 

the maximum number of \textbf{eigenstates} to compute, in terms of both the principal and angular quantum numbers;
\item 

the numerical \textbf{grid} parameters, in particular the grid size;
\item 

the \textbf{convergence} tolerances, Eqs. (14) to (16);
\item 

the \textbf{SCF} parameters, i.e. the mixing fraction and the maximum number of iterations.\end{itemize}

The first three items in this list essentially control the accuracy of the calculation.
In principle, for each SCF calculation — i.e. a unique set of physical and model inputs — these parameters should be independently varied until some property (such as the total free energy) is considered suitably converged with respect to that parameter.
Changing the SCF parameters should not affect the final results (within the convergence tolerances), only the number of iterations in the SCF cycle.

Let us now consider an example SCF calculation, using the \texttt{\DUrole{code}{Atom}} and \texttt{\DUrole{code}{model}} objects we have already defined:\vspace{1mm}
\begin{Verbatim}[commandchars=\\\{\},fontsize=\footnotesize]
\PY{k+kn}{from} \PY{n+nn}{atoMEC} \PY{k+kn}{import} \PY{n}{config}
\PY{n}{config}\PY{o}{.}\PY{n}{numcores} \PY{o}{=} \PY{o}{\PYZhy{}}\PY{l+m+mi}{1} \PY{c+c1}{\PYZsh{} parallelize}

\PY{n}{nmax} \PY{o}{=} \PY{l+m+mi}{3} \PY{c+c1}{\PYZsh{} max value of principal quantum number}
\PY{n}{lmax} \PY{o}{=} \PY{l+m+mi}{3} \PY{c+c1}{\PYZsh{} max value of angular quantum number}

\PY{c+c1}{\PYZsh{} run SCF calculation}
\PY{n}{scf\PYZus{}out} \PY{o}{=} \PY{n}{model}\PY{o}{.}\PY{n}{CalcEnergy}\PY{p}{(}
 \PY{n}{nmax}\PY{p}{,}
 \PY{n}{lmax}\PY{p}{,}
 \PY{n}{grid\PYZus{}params}\PY{o}{=}\PY{p}{\PYZob{}}\PY{l+s+s2}{\PYZdq{}}\PY{l+s+s2}{ngrid}\PY{l+s+s2}{\PYZdq{}}\PY{p}{:} \PY{l+m+mi}{1500}\PY{p}{\PYZcb{}}\PY{p}{,}
 \PY{n}{scf\PYZus{}params}\PY{o}{=}\PY{p}{\PYZob{}}\PY{l+s+s2}{\PYZdq{}}\PY{l+s+s2}{mixfrac}\PY{l+s+s2}{\PYZdq{}}\PY{p}{:} \PY{l+m+mf}{0.7}\PY{p}{\PYZcb{}}\PY{p}{,}
 \PY{p}{)}
\end{Verbatim}
\vspace{1mm}
We see that the first two parameters passed to the \texttt{\DUrole{code}{CalcEnergy}} function are the \texttt{\DUrole{code}{nmax}} and \texttt{\DUrole{code}{lmax}} quantum numbers, which specify the number of eigenstates to compute.
Precisely speaking, there is a unique Hamiltonian for each value of the angular quantum number $l$ (and in a spin-polarized calculation, also for each spin quantum number).
The sparse diagonalization routine then computes the first \texttt{\DUrole{code}{nmax}} eigenvalues for each Hamiltonian.
In atoMEC, these diagonalizations can be run in parallel since they are independent for each value of $l$.
This is done by setting the \texttt{\DUrole{code}{config.numcores}} variable to the number of cores desired (\texttt{\DUrole{code}{config.numcores=-1}} uses all the available cores) and handled via the joblib library \DUrole{cite}{joblib}.

The remaining parameters passed to the \texttt{\DUrole{code}{CalcEnergy}} function are optional; in the above, we have specified a grid size of 1500 points and a mixing fraction $\alpha=0.7$.
The above code automatically prints the output seen in Fig. 5.
This output shows the SCF cycle and, upon completion, the breakdown of the total free energy into its various components, as well as other useful information such as the KS energy levels and their occupations.\begin{figure}[]\noindent\makebox[\columnwidth][c]{\includegraphics[width=\columnwidth]{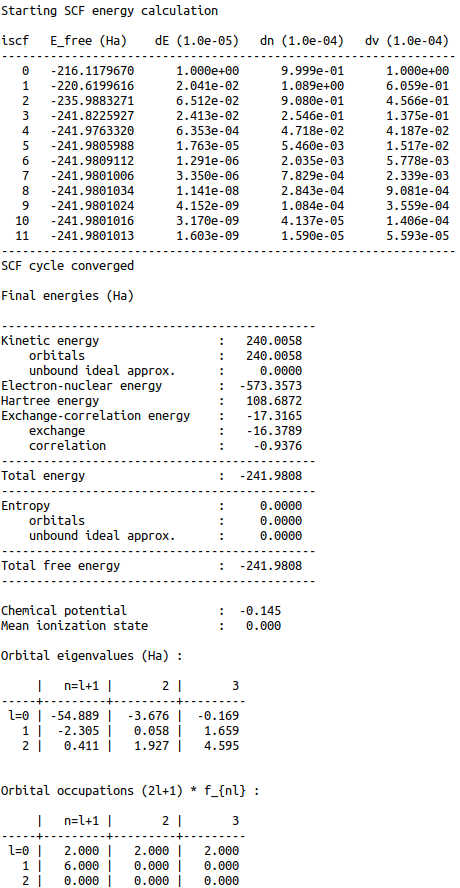}}
\caption{Auto-generated print statement from calling the \texttt{\DUrole{code}{ISModel.CalcEnergy}} function}
\end{figure}

Additionally, the output of the SCF function is a dictionary containing the \texttt{\DUrole{code}{staticKS.Orbitals}}, \texttt{\DUrole{code}{staticKS.Density}}, \texttt{\DUrole{code}{staticKS.Potential}} and \texttt{\DUrole{code}{staticKS.Density}} objects.
For example, one could extract the eigenfunctions as follows:\vspace{1mm}
\begin{Verbatim}[commandchars=\\\{\},fontsize=\footnotesize]
\PY{n}{orbs} \PY{o}{=} \PY{n}{scf\PYZus{}out}\PY{p}{[}\PY{l+s+s2}{\PYZdq{}}\PY{l+s+s2}{orbitals}\PY{l+s+s2}{\PYZdq{}}\PY{p}{]} \PY{c+c1}{\PYZsh{} orbs object}
\PY{n}{ks\PYZus{}eigfuncs} \PY{o}{=} \PY{n}{orbs}\PY{o}{.}\PY{n}{eigfuncs} \PY{c+c1}{\PYZsh{} eigenfunctions}
\end{Verbatim}
\vspace{1mm}
The initialization of the SCF procedure is shown in the third and fourth rows of Fig. 2, with the SCF procedure itself shown in the remaining rows.

This completes the section on the code structure and algorithmic details.
As discussed, with the output of an SCF calculation, there are various kinds of post-processing one can perform to obtain other properties of interest.
So far in atoMEC, these are limited to the computation of the pressure (\texttt{\DUrole{code}{ISModel.CalcPressure}}), the electron localization function (\texttt{\DUrole{code}{atoMEC.postprocess.ELFTools}}) and the Kubo–Greenwood conductivity (\texttt{\DUrole{code}{atoMEC.postprocess.conductivity}}).
We refer readers to our pre-print \DUrole{cite}{arxiv_KG} for details on how the electron localization function and the Kubo–Greenwood conductivity can be used to improve predictions of the mean ionization state.

\subsection{Case-study: Helium%
  \label{case-study-helium}%
}

In this section, we consider an application of atoMEC in the WDM regime.
Helium is the second most abundant element in the universe (after hydrogen) and therefore understanding its behavior under a wide range of conditions is important for our understanding of many astrophysical processes.
Of particular interest are the conditions under which helium is expected to undergo a transition from insulating to metallic behavior in the outer layers of white dwarfs, which are characterized by densities of around $1-20 \textrm{ g cm}^{-3}$ and temperatures of $10-50$ kK \DUrole{cite}{Hellium_metal}.
These conditions are a typical example of the WDM regime.
Besides predicting the point at which the insulator-to-metallic transition occurs in the density-temperature spectrum, other properties of interest include equation-of-state data (relating pressure, density, and temperature) and electrical conductivity.

To calculate the insulator-to-metallic transition point, the key quantity is the electronic \emph{band-gap}.
The concept of band-structures is a complicated topic, which we try to briefly describe in layman's terms.
In solids, electrons can occupy certain energy ranges — we call these the energy bands.
In insulating materials, there is a gap between these energy ranges that electrons are forbidden from occupying — this is the so-called band-gap.
In conducting materials, there is no such gap, and therefore electrons can conduct electricity because they can be excited into any part of the energy spectrum.
Therefore, a simple method to determine the insulator-to-metallic transition is to determine the density at which the band-gap becomes zero.

In Fig. 6, we plot the density-of-states (DOS) as a function of energy, for different densities and at fixed temperature $\tau=50$ kK.
The DOS shows the energy ranges that the electrons are allowed to occupy; we also show the actual energies occupied by the electrons (according to Fermi–Dirac statistics) with the black dots.
We can clearly see in this figure that the band-gap (the region where the DOS is zero) becomes smaller as a function of density.
From this figure, it seems the transition from insulating to metallic state happens somewhere between 5 and 6 $\textrm{g cm}^{-3}.$\begin{figure}[]\noindent\makebox[\columnwidth][c]{\includegraphics[scale=1.00]{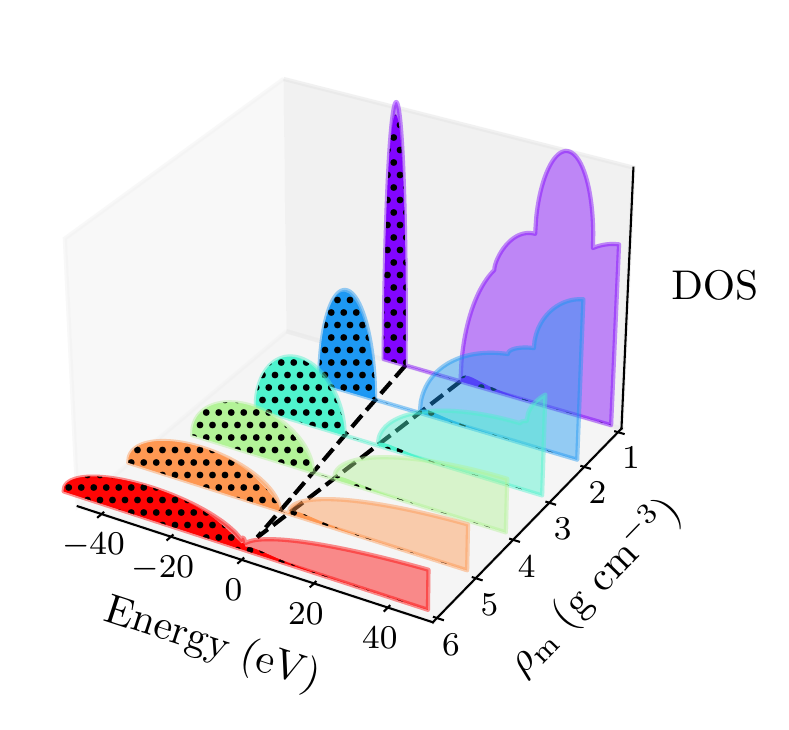}}
\caption{Helium density-of-states (DOS) as a function of energy, for different mass densities $\rho_\textrm{m}$, and at temperature $\tau=50$ kK.
Black dots indicate the occupations of the electrons in the permitted energy ranges.
Dashed black lines indicate the band-gap (the energy gap between the insulating and conducting bands).
Between 5 and 6 $\textrm{g cm}^{-3}$, the band-gap disappears.}
\end{figure}

In Fig. 7, we plot the band-gap as a function of density, for a fixed temperature $\tau=50$ kK.
Visually, it appears that the relationship between band-gap and density is linear at this temperature.
This is confirmed using a linear fit, which has a coefficient of determination value of almost exactly one, $R^2=0.9997$.
Using this fit, the band-gap is predicted to close at $5.5\ \textrm{g cm}^{-3}$.
Also in this figure, we show the fraction of ionized electrons, which is given by $\bar{Z}/N_\textrm{e}$, using Eq. (\DUrole{ref}{eq:MIS}) to calculate $\bar{Z}$, and $N_\textrm{e}$ being the total electron number.
The ionization fraction also relates to the conductivity of the material, because ionized electrons are not bound to any nuclei and therefore free to conduct electricity.
We see that the ionization fraction mostly increases with density (excepting some strange behavior around $\rho_\textrm{m}=1\ \textrm{g cm}^{-3}$), which is further evidence of the transition from insulating to conducting behaviour with increasing density.\begin{figure}[]\noindent\makebox[\columnwidth][c]{\includegraphics[scale=1.00]{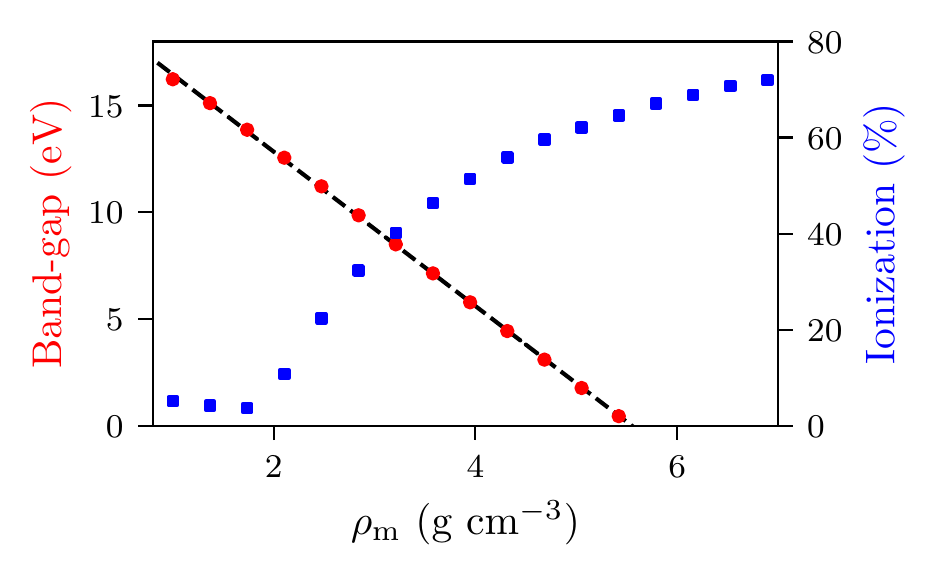}}
\caption{Band-gap (red circles) and ionization fraction (blue squares) for helium as a function of mass density, at temperature $\tau=50$ kK.
The relationship between the band-gap and the density appears to be linear.}
\end{figure}

As a final analysis, we plot the pressure as a function of mass density and temperature in Fig. 8.
The pressure is given by the sum of two terms: (i) the electronic pressure, calculated using the method described in Ref. \DUrole{cite}{AA_pressure}, and (ii) the ionic pressure, calculated using the ideal gas law.
We observe that the pressure increases with both density and temperature, which is the expected behavior.
Under these conditions, the density dependence is much stronger, especially for higher densities.

The code required to generate the above results and plots can be found in \href{https://github.com/atomec-project/Helium-white-dwarfs}{this repository}.\begin{figure}[]\noindent\makebox[\columnwidth][c]{\includegraphics[scale=1.00]{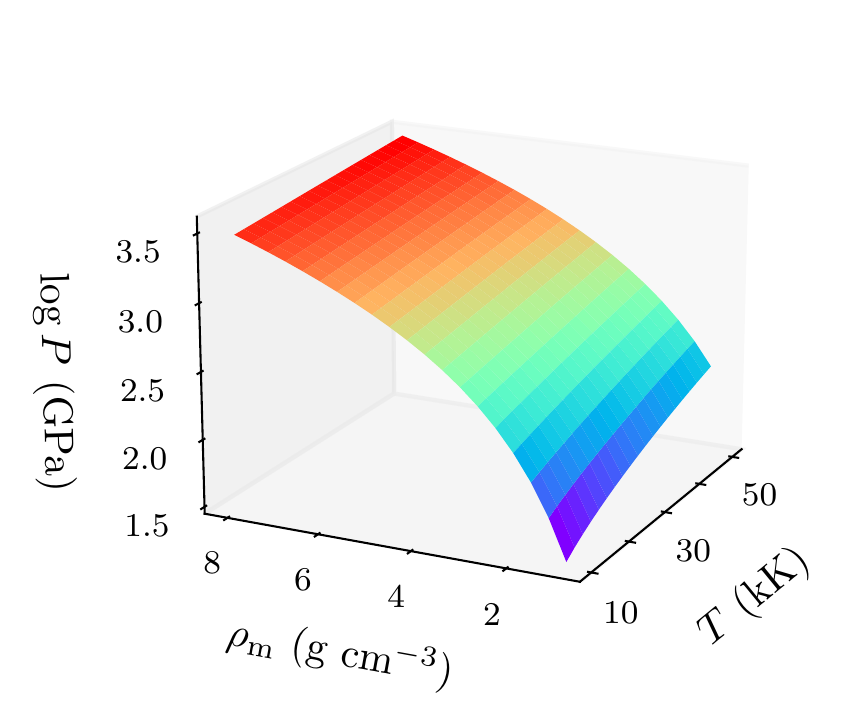}}
\caption{Helium pressure (logarithmic scale) as a function of mass density and temperature.
The pressure increases with density and temperature (as expected), with a stronger dependence on density.}
\end{figure}

\subsection{Conclusions and future work%
  \label{conclusions-and-future-work}%
}

In this paper, we have presented atoMEC: an  average-atom Python code for studying materials under extreme conditions.
The open-source nature of atoMEC, and the choice to use (pure) Python as the programming language, is designed to improve the accessibility of average-atom models.

We gave significant attention to the code structure in this paper, and tried as much as possible to connect the functions and objects in the code with the underlying theory.
We hope that this not only improves atoMEC from a user perspective, but also facilitates new contributions from the wider average-atom, WDM and scientific Python communities.
Another aim of the paper was to communicate how atoMEC benefits from a strong ecosystem of open-source scientific libraries — especially the Python libraries NumPy, SciPy, joblib and mendeleev, as well as LIBXC.

We finish this paper by emphasizing that atoMEC is still in the early stages of development, and there are many opportunities to improve and extend the code.
These include, for example:
\begin{itemize}
\item 

adding new average-atom models, and different approximations to the existing \texttt{\DUrole{code}{models.ISModel}} model;
\item 

optimizing the code, in particular the routines in the \texttt{\DUrole{code}{numerov}} module;
\item 

adding new postprocessing functionality, for example to compute structure factors;
\item 

improving the structure and design choices of the code.\end{itemize}

Of course, these are just a snapshot of the avenues for future development in atoMEC.
We are open to contributions in these areas and many more besides.

\subsection{Acknowledgements%
  \label{acknowledgements}%
}

This work was partly funded by the Center for Advanced
Systems Understanding (CASUS) which is financed by
Germany’s Federal Ministry of Education and Research
(BMBF) and by the Saxon Ministry for Science, Culture
and Tourism (SMWK) with tax funds on the basis of the
budget approved by the Saxon State Parliament.
\bibliographystyle{alphaurl}
\bibliography{main}

\end{document}